\PassOptionsToPackage{hyphens}{url}
\documentclass[10pt, conference, compsocconf]{IEEEtran}

\usepackage{latexsym, amsmath, amsfonts, amsthm, bm, bbm, nicefrac, stmaryrd, thmtools, enumitem, centernot, wasysym}
\usepackage{url, graphicx, xcolor, makecell, rotating, colortbl}
\definecolor{svlinks}{rgb}{.0,0.3,0.6}
\usepackage[hang]{footmisc}
\usepackage[bookmarks,colorlinks=true,pdfhighlight=/O,linkcolor=svlinks,urlcolor=svlinks,citecolor=svlinks,
           pdftitle={Small Towns, Big Questions: Methodological Insights into Use Case Selection for Digital Twins in Small Towns},
           pdfauthor={Lucy Temple, Laura Kaltenbrunner, Gabriela Viale Pereira, Lukas Daniel Klausner},
           pdfsubject={},
           pdfkeywords={}]{hyperref}
\usepackage[T2A,T1]{fontenc}
\usepackage[utf8]{inputenc}
\usepackage{newunicodechar}
\usepackage[russian,greek,UKenglish]{babel}

\usepackage{epsfig, listings}
\usepackage{multirow}
\usepackage{float}
\usepackage[section]{placeins}
\usepackage[all]{nowidow}
\usepackage{fnpct}
\usepackage{soul}

\usepackage{tikz, wrapfig, caption, xhfill}
\usetikzlibrary{matrix, backgrounds, positioning}
\usepackage[normalem]{ulem}
\usepackage{booktabs}
\usepackage{array}
\usepackage[subfolder]{gnuplottex}
\usepackage{xfrac}
\usepackage{float}

\usepackage{placeins}
\usepackage[all]{nowidow}
\usepackage{fnpct}
\usepackage[capitalise,noabbrev]{cleveref}

\setlist{nosep}

\usepackage{etoolbox}
\pretocmd{\eqref}{Eq.~}{}{}

\DeclareRobustCommand{\okina}{%
  \raisebox{\dimexpr\fontcharht\font`A-\height}{%
    \scalebox{0.8}{`}%
  }%
}
\newunicodechar{ʻ}{\okina}

\newcommand*{\TableHigh}{\CIRCLE}
\newcommand*{\TableMedium}{\RIGHTcircle}
\newcommand*{\TableLow}{\Circle}
\newcommand*{\TableYes}{\CIRCLE}
\newcommand*{\TableSimplified}{$\sim$}
\newcommand*{\TableNo}{\Circle}
\newcommand*{\TablePossible}{?}
\newcommand*{\TableNecessary}{!}

\begin{document}

\title{Small Towns, Big Questions: Methodological Insights into Use Case Selection for Digital~Twins~in Small Towns}

\author{
\IEEEauthorblockN{Lucy Temple, Gabriela Viale Pereira}
\IEEEauthorblockA{
{\it Department for E-Governance and Administration}\\
{\it University for Continuing Education Krems}\\
Krems, Austria\\
\{lucy.temple, gabriela.viale-pereira\}@donau-uni.ac.at}
\and
\IEEEauthorblockN{Laura Kaltenbrunner, Lukas Daniel Klausner}
\IEEEauthorblockA{
{\it Institute of IT Security Research}\\
{\it St.\ P\"olten UAS}\\
St.\ P\"olten, Austria\\
\{laura.kaltenbrunner, lukas.daniel.klausner\}@fhstp.ac.at}
}




\maketitle

\begin{abstract}
    Selecting appropriate use cases for implementing digital solutions in small towns is a recurring challenge for smart city projects. This paper presents a transdisciplinary methodology for systematically identifying and evaluating such use cases, drawing from diverse academic disciplines and practical expertise. The proposed methodology was developed and implemented in Lower Austria, with a particular focus on the small towns that are characteristic of a region lacking major urban centres. Through semi-structured interviews and collaborative workshops (e.\,g.\ a needs requirements workshop) with various relevant stakeholders, fifteen possible use cases were first identified. Then these use cases were categorised and assessed based on criteria such as feasibility, usefulness, the need for biological or human modelling, and overall complexity. Based on these characteristics, three use cases were selected for further development. These will be the basis of digital twin solutions for supporting decision-making and public outreach regarding policy decisions in those fields. Our proposed methodology emphasises stakeholder engagement to ensure robust data collection and alignment with practical requirements and the involved towns’ current needs. We thus provide a replicable framework for researchers and practitioners aiming to implement digital twin tools in future smart city initiatives in non-urban and rural contexts.
\end{abstract}


\section{Introduction}
\label{sec:intro}
Technological advancements are increasingly being integrated into urban environments worldwide, coinciding with a surge in funding and prominence for smart city policies~\cite{CaragliuDelBo2019}. The use of digital tools to manage societal challenges has promoted a change in the ways public administrations work, but also has an impact on the governance approaches~\cite{MergelEdelmann2019}, because cities are complex systems that integrate a variety of actors, elements and networks. As the digital transformation of cities accelerates, it becomes essential to integrate stakeholders and identify local needs, recognising that a one-size-fits-all approach is insufficient.

\begin{figure*}[t]
    \begin{minipage}{\textwidth}
    \centering
    \begin{tikzpicture}[scale=1]
	\tikzstyle{every node}=[font=\small]
	\tikzstyle{block} = [rectangle, draw, thick, text width=12.5em, text centered, minimum height=1em]
        
        \path (0,0) node(Block1) [block] {Data collection through twelve interviews};
	\node (Block2) [block, right = 3em of Block1] {Identification of fifteen initial use cases};
	\node (Block3) [block, below = 3em of Block2] {Internal use case prioritisation resulting in three use cases};
	\node (Block4) [block, left = 3em of Block3] {Data and requirements gathering workshop};
        
	\draw[thick,->,shorten >= 6pt, shorten <= 6pt] (Block1.east) -- (Block2.west);
	\draw[thick,->,shorten >= 6pt, shorten <= 6pt] (Block2.south) -- (Block3.north);
	\draw[thick,->,shorten >= 6pt, shorten <= 6pt] (Block3.west) -- (Block4.east);
    \end{tikzpicture}
    \smallskip
    \caption{Overview of the four steps of the proposed methodology in our study, corresponding to the four subsections of \autoref{sec:methodology}.}
    \label{fig:methodologysteps}
    \end{minipage}
\end{figure*}
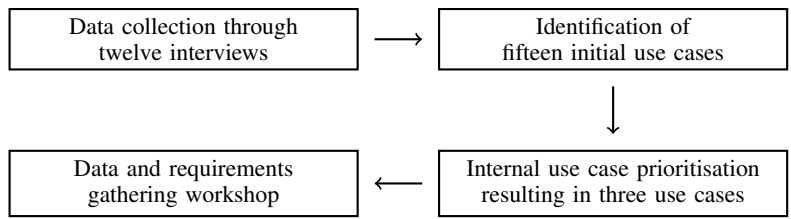

Transdisciplinary processes are particularly well-suited for addressing complex, real-world challenges faced by small towns, especially when selecting and implementing use cases that require a holistic understanding of local contexts. These processes are tackling problems in an interdisciplinary way, extending beyond academic stakeholders, actively engaging practitioners, policymakers and society as beneficiaries, and enhancing the relevance of solutions~\cite{EiblTemple2022}. For selecting use cases in small towns, we adapt the methodology proposed by Scholz~\cite[Fig.\ 1]{Scholz2020}, which emphasises the need for collaboration at the science–practice–policy interface to address multidimensional and multifaceted challenges, such as the adoption of digital twin solutions in cities. As highlighted by Almulhim et al.~\cite{AlmulhimSharifi2024}, creating inclusive, safe and climate-resilient urban environments requires collective efforts that actively involve policymakers, academics, practitioners and citizens. In this context, transdisciplinary approaches bridge the gap between specific subject knowledge of scientists in theories, methodologies and tools and the local knowledge of practitioners and policymakers, who understand the subtle details and applicable solutions within their communities. These processes also facilitate societal learning, enabling stakeholders to co-develop socially robust orientations for sustainable development~\cite{ScholzSteiner2015}, referring to solutions that not only incorporate scientific state-of-the-art knowledge but also integrate local experiences and practical wisdom by acknowledging uncertainty and knowledge gaps as well as ensuring transparency about the constraints of the study~\cite{Scholz2011}. By following these approaches in the context of small towns, transdisciplinary processes support the selection of cases that are context-sensitive, feasible and aligned with communities' goals and needs for sustainable development.

A digital twin is the virtual representation of a physical system, alongside its environment and processes, through the exchange of information between the physical and virtual systems~\cite{VanDerHornMahadevan2021}. The goal of digital twins is to simulate the behaviour of an object and enable decision-making based on predictions~\cite{ShiPan2023}. Cities and towns around the world are creating vast amounts of data that, without appropriate tools and technologies, will exceed human capabilities for analysis and prediction. Digital twins for cities can be used to model policy decision~\cite{AlamElSaddik2017}; they can be created by using the data collected with the smart technologies in cities~\cite{MohammadiVimal2020}.

While digital twins have proven valuable in urban planning and smart city initiatives, their adoption in small towns remains a challenge due to resource constraints, governance structures and diverse local needs. The objective of this study is to explore how transdisciplinary methods can be applied to identify and assess the suitability of use cases for developing digital twin solutions in small towns. By integrating insights from multiple disciplines and engaging key stakeholders, including policymakers, researchers and local communities, we aim to establish a replicable framework that supports future research to identify use cases for digital solutions in small-town contexts.

\section{Methodology}
\label{sec:methodology}
To explore societal challenges faced by municipalities that could serve as use cases for a digital twin, as well as to identify the requirements for its development, we designed a multi-stage qualitative approach (see \autoref{fig:methodologysteps}).

First, guided interviews with political stakeholders (such as mayors, district governors and state government personnel) were performed to gather information about societal challenges within municipalities in Lower Austria. Furthermore, these interviews served as a way of identifying additional relevant participants for the interviews as well as the later data and requirements gathering workshop. The information gathered through the interviews was then consolidated and transformed into specific use cases that reflected the societal challenges addressed. All use cases were ranked by a set of criteria described in more detail in \autoref{subsec:usecaseprioritisation}. The highest-ranked use cases were then selected and further discussed in a data and requirements gathering workshop with previous interview partners. Here, the goal was to validate the use cases, gather more information about them and identify specific requirements and options for building and handling a corresponding digital twin (e.\,g.\ availability of data).

\begin{table*}[th!]
    \begin{minipage}{\textwidth}
    \centering
    \begin{tabular}{p{12cm}ll} \toprule
    \bf Position & \bf Administrative division & \bf Way of contact \\ \midrule
    Head of Unit: Hydrographic Service and Flood Forecasts & state government & existing contact \\
    Management: Department of Environmental and Plant Law/Head of the Business, Sport and Tourism Group & state government & existing contact \\
    General Road Services Department & state government & existing contact \\
    City Marketing & municipality & existing contact \\
    Mayor & municipality & snowball method \\
    District Governor & district & snowball method \\
    General Road Services Department Strategic Controlling Department & state government & snowball method \\
    Road Operations Department Winter Road Maintenance and Traffic Management Department & state government & snowball method \\
    Air Quality Monitoring Network Department of Environmental and Plant Engineering & state government & snowball method \\
    Head of the Environmental Coordination Department Environmental and Plant Engineering & state government & snowball method \\
    Manager of the Model Region Unteres Traisental and Fladnitztal & municipality & snowball method \\
    City Council – Member of the Tourism and Economy Committee & muncipality & snowball method \\ \bottomrule
    \end{tabular}
    \caption{Overview of interview participants, including their official position title, the level of government they are active in and our way of contacting them (either existing working relationships or through a snowball approach based on these initial contacts).}
    \label{tab:interviewparticipants}
    \end{minipage}
\end{table*}

\subsection{Data Collection through Interviews}
\label{subsec:datacollection}
To identify societal challenges within municipalities from the perspective of political decision-makers, government officials in Lower Austria were selected as the target group for both the guided interviews and the following step of the needs assessment workshops. For the interview sample, the focus was on people within the governance sector that are specialised in one of the fields identified as the most relevant for applying digital twin solutions in the smaller municipalities. These fields were identified through a literature review process (cf.~\cite{VialePereiraTemple2024}): environment, governance and policy-making, society and ICT, and mobility and transport. (In practical terms, this means domains such as tourism, energy, sustainability, agriculture, waste management, risk control, communication and traffic management.)

However, since we considered it unlikely that government officials in Lower Austria could be engaged and convinced to participate in this study through a cold call invitation, a snowball approach (wherein researchers can access new interviewee partners through contact information provided by previous interviewees) was used to increase the interview sample~\cite{Noy2008}. Although this approach does not generate a random sample, it allows access to a very specific community that is difficult to reach or identify beforehand~\cite[p.\ 413]{Akremi2022}). While for our purposes, this kind of availability sampling sufficed, it runs the risk of suffering from methodological bias. For less exploratory and/or more complex configurations of possible stakeholders, a more representative sampling method would be advisable, possibly in conjunction with applying power/interest matrices~\cite[p.\ 156 ff.]{JohnsonScholes2008} as an analytical tool for understanding and managing different types of stakeholder relationships.

Starting with four government officials with whom there was already an existing working relationship through previous research projects, a total of twelve experts were identified, and interviewed; after the twelfth interview, saturation was reached, as no new insights or information were contributed~\cite{WellerVickers2018}. The twelve experts were individually and sequentially contacted by email and invited to an in-person meeting between June and July 2024. Here, they were also informed about the project, its goals and the potential outcomes for the municipality as well as about the scope and modality of the interview. \autoref{tab:interviewparticipants} presents a detailed account of the interview partners.

Semi-structured guided interviews as part of qualitative research were selected as the first methodological approach, as they allow focussing on the experience of a person and their view on the research topic and are therefore a suitable choice for gathering a base of information where there has not been one before~\cite[p.\ 122]{Schumann2018}. Furthermore, in semi-structured interviews, the interviewees are motivated to think freely, generating a broader overview on the research topic. In the case of this research, the focus was on the experience of government officials with the occurrence and handling of societal challenges in municipalities in Lower Austria. The goal of the interview guide (available in the appendix in \autoref{anx:interviewguide}) was therefore to identify specific societal challenges within Lower Austrian municipalities that are present within the focus area of the interviewee, availability of corresponding data, and the identification of further interview partners. If additional societal challenges emerged during an interview that the interviewee was not deeply familiar with, another knowledgeable individual from their network was identified and contacted, thereby expanding the sample size. The guided interviews were previously tested and had an expected duration of 45~minutes. Every interview was recorded and transcribed using Whisper, an automatic speech recognition system developed by OpenAI.

\subsection{Use Case Identification}
\label{subsec:usecaseidentification}
Based on the interview transcripts, a deductive qualitative content analysis following Mayring's~\cite{Mayring2022} methodology was conducted. This approach allows for systematic categorisation and analysis of interview data based on predefined categories relevant to the research question. The content analysis aimed at identifying societal challenges described by the participants and transforming these challenges into tangible use cases for government decision-making, particularly in the context of digital twin development. To facilitate a structured analysis, the following predefined categories were established:
\begin{itemize}
    \item General description of the societal challenge: What challenge has the interview partner been describing?
    \item Occurrence of the societal challenge: Where does this challenge occur? How large is the impact area of the challenge?
    \item Described urgency of the societal challenge: How urgent is this challenge according to the interview partner? Was it prioritised?
    \item Available data: Which data were mentioned to be useful for this use case and are they available through this interview partner?
    \item Additionally needed data sets: Which data sets were mentioned to be useful in handling this challenge but are not available through this interview partner?
    \item Sustainability: What impact does the challenge have on sustainability and which effects can come from helping government officials handle this challenge by building a digital twin?
    \item Governance: Which decisions are connected to this challenge? Where could a digital twin help government officials make policy decisions?
    \item Needs to be evaluated: What information does not seem fully complete or needs further evaluation in the data and requirements gathering workshop?
    \item Final thoughts: How can this challenge be summarised?
\end{itemize}

Each interview was analysed individually according to these categories, ensuring a consistent approach to identifying the most critical societal challenges across municipalities. The data were then consolidated to identify recurring themes and patterns. The goal of the consolidation was to find challenges that were described more often in the different interviews or that could be combined into one use case. Based on this consolidation, fifteen possible use cases were identified. These are described below in \autoref{subsec:initialusecases}.

\begin{figure*}[!ht]
    \begin{minipage}{\textwidth}
    \centering
    \includegraphics[height=8.5cm]{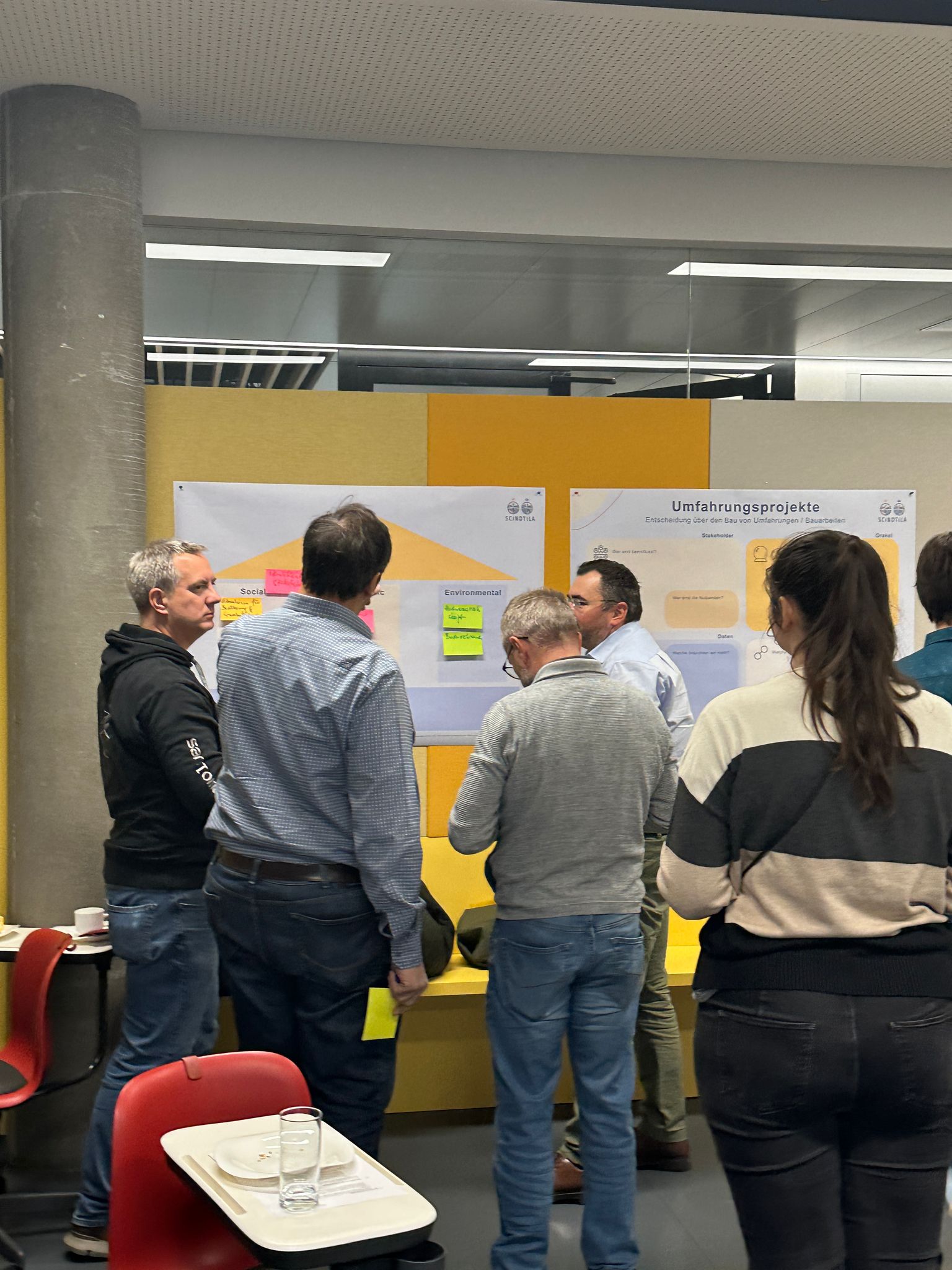}
    \includegraphics[height=8.5cm]{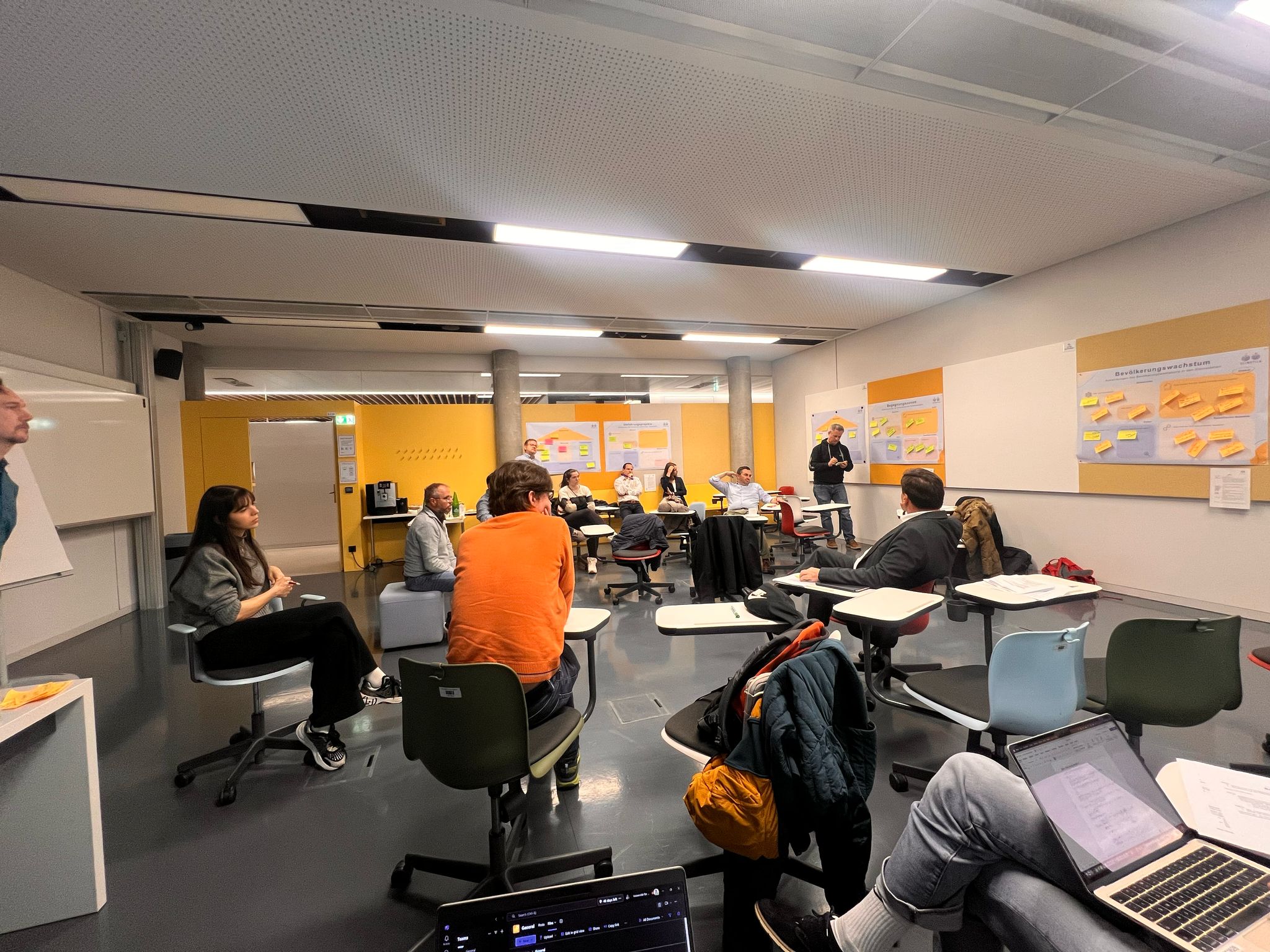}
    \caption{Impressions from the data and requirements gathering workshop held at St.\ Pölten University of Applied Sciences, as described in \autoref{subsec:workshop}.}
    \label{fig:workshopimpressions}
    \end{minipage}
\end{figure*}

\subsection{Internal Use Case Prioritisation}
\label{subsec:usecaseprioritisation}
These fifteen use cases were subsequently evaluated in a dedicated internal workshop to select the most promising use cases to develop further. In a first step, a set of criteria was chosen as the foundation of the selection process, with a focus on considering the various implementation requirements, fields of expertise and data sources required to further develop the use cases. To derive the criteria, we drew upon team-internal research experience in e-government, science and technology studies, and computer science. The first-hand experience of the research team provided a clear understanding of the technological possibilities of the available software as well as the limitations associated with digital twin simulations. Moreover, the collective knowledge on the various topics as well as the lack of in-depth expertise in biological systems were taken into account to define and evaluate the criteria.\footnote{Biological modelling presents a significant challenge in the context of digital twins~\cite{TrantasPlug2023}. There are two main limitations in this context – on the one hand, the scarcity of high-resolution real-time biological data, on the other hand, the complex non-linear interdependencies between various biological variables such as temperature, humidity, soil composition and water flow. Adequately integrating such biological modelling demands not only interoperability between diverse data sources, but also domain-specific calibration, requiring a significant degree of expert knowledge not available to the researchers involved.} The chosen criteria were as follows: 
\begin{itemize}
    \item Usefulness (low/medium/high): What would the potential impact of this use case be in practice? The evaluation for this criterion was based both on the described and perceived urgency of the use case by the involved stakeholders, the number of municipalities or people affected by the issue (directly and indirectly) as well as our team’s estimation of the added value of the realisation of this use case when compared to existing approaches. 
    \item Human behaviour modelling required (no/simplified/yes): To what degree does this use case involve the modelling of aspects of human behaviour (e.\,g.\ traffic, commerce, leisure activities), and how difficult would this aspect be in the team’s estimation? 
    \item Biological system modelling required (no/yes): Does this use case involve any kind of modelling of biological systems (e.\,g.\ animal populations, weather/climate conditions, water flows, plant growth)? 
    \item High degree of required domain knowledge (no/simplified/yes): Is there a high degree of domain knowledge required to be able to implement and interact with this use case? 
    \item Geographical modelling (optional/necessary): Is it absolutely necessary to apply geographical modelling in this use case, or just optional? 
    \item 3D modelling (no/optional/necessary): Is there any involvement of 3D modelling in this use case, and if so, is it necessary or just optional? 
    \item Overall complexity (low/medium/high): Considering the previous criteria, what is the overall complexity of this use case? 
\end{itemize}
In addition to these criteria, the impact on governance and sustainability was also considered for all use cases (but not ranked on a scale). Based on these criteria, three use cases emerged as most promising and feasible in light of their respective high usefulness and manageable overall complexity (see \autoref{subsec:summarytable} for details); one of the three was the natural combination of two previously separate use cases.

\begin{table*}[th!]
    \begin{minipage}{\textwidth}
    \centering
    \begin{tabular}{cp{2.75cm}p{13.5cm}} \toprule
    \bf No. & \bf Short title & \bf Description \\ \midrule
    1 & Opening hours & Which criteria should be considered when evaluating the adjustment of shop opening hours in cities and larger municipalities in Lower Austria, particularly regarding economic factors, health aspects and energy consumption? \\
    2 & Blackouts and increased energy consumption & Which aspects should be included in municipal crisis plans for blackouts in Lower Austria and how can simulations (e.g. digital twins) help identify necessary adaptations? \\
    3 & Population growth & How will a projected population increase of 21,000 people in Baden over the next 15 years impact critical infrastructure such as traffic, housing, childcare, elderly care and water supply, and what measures should be planned? \\
    4 & Wastewater & What are the effects of increased wastewater discharge and elevated water temperatures on the flora and fauna of the Mühlbach and how can these impacts be mitigated? \\
    5 & Medical supply & How can establishing optimally located medical centres reduce hospital strain caused by an ageing and growing population and what factors should be considered in determining their placement? \\
    6 & Decluttering of traffic signs & How does the decluttering of traffic signs in municipalities in Lower Austria impact traffic flow and safety and what effects arise from removing individual signs versus combinations of signs? \\
    7 & Winter service management & How can winter services in Lower Austria, including road cleaning and salt scattering, be optimised using the current vehicle fleet, weather forecasts and additional vehicles from contractors, while considering the different types of vehicles and road conditions? \\
    8 & Added value through bypass projects & How can a digital twin be used to visualise the benefits of bypass projects, like those proposed for Wiener Neustadt, to help garner public support and address concerns about the removal of community spaces such as parks? \\
    9 & Traffic flow planning & How can construction site planning and event scheduling be optimised to minimise disruption to traffic flow and how can visualisations help explain the rationale behind the timing and planning of such projects to citizens? \\
    10 & Drainage management and sedimentation planning & How can a digital twin be used to model scenarios for settlement planning, considering the impact of extreme weather events like heavy rainfall on drainage, and what factors should be considered in optimising drainage and reducing flooding risks? \\
    11 & Climate change effects on fish populations & How does climate change, particularly variations in precipitation, peak events and dry periods, impact fish populations and their health and what measures can be taken to monitor and mitigate these effects over time? \\
    12 & Climate change and mobility & How does climate change, with rising temperatures and shifting weather patterns, impact commuting behaviour and which modes of transportation should be prioritised for investment to adapt to these changes? \\
    13 & Climate change effects on city planning & How does climate change, particularly hotter summers and extreme weather conditions, affect the liveability of urbanised municipalities and what strategies can be implemented to enhance shaded areas and green spaces for increased resilience and comfort? \\
    14 & Shared spaces & How does converting traffic roads into shared spaces or pedestrian zones impact mobility patterns, alternative transportation use and inner-city shopping and economic activity and what are the potential benefits and challenges of such transformations? \\
    15 & Climate change effects on wine & How will climate change, with rising temperatures and reduced rainfall, affect wine regions in Lower Austria, particularly in terms of harvest times and wine quality, and what strategies can be implemented to adapt to these changes? \\ \bottomrule
    \end{tabular}
    \caption{Overview of initial use cases with short descriptions of the guiding questions and key challenges.}
    \label{tab:usecaseoverview}
    \end{minipage}
\end{table*}

\subsection{Data and Requirements Gathering Workshop}
\label{subsec:workshop}
To refine the requirements for the three selected use cases, a data and requirement gathering workshop was conducted (cf.\ \autoref{fig:workshopimpressions}). There were a total of twelve external participants and four workshop organisers. Most participants had previously taken part in the interview phase (cf.\ \autoref{subsec:datacollection}), which allowed for a deeper exploration of the selected use case by building on prior discussions. This continuity enabled detailed insights and reflections on the challenges and opportunities associated with each case. Some interview participants whose proposed use cases were not selected still contributed valuable perspectives, offering fresh insights and constructive critiques that enriched the overall analysis.

The primary goal of the workshop was to identify the specific requirements for digital twins across three selected use cases. This included assessing the societal and governance impacts, defining the questions the digital twins should address, determining data needs, and understanding the necessary level of granularity for decision-making. The workshop followed a structured agenda designed to facilitate collaborative discussions and iterative knowledge building. To begin with, an overview of the project goals was given, followed by a detailed presentation of the three selected use cases. Participants had to evaluate the sustainability impact, stakeholder identification, potential end users, and key questions for the digital twin to answer. A particularly important topic discussed in the sessions was the data requirements for the digital twin.

\section{Results}
\label{sec:results}
This section presents key findings derived from the proposed methodology. First, in \autoref{subsec:initialusecases} the fifteen initial use cases are presented; these were identified through deductive qualitative content analysis based on the interviews described in \autoref{subsec:datacollection} and \ref{subsec:usecaseidentification}. Next, \autoref{subsec:summarytable} presents a table summarising the results of the internal use case prioritisation. Finally, \autoref{subsec:selectedusecases} describes the three selected use cases in detail. 

\subsection{Initial Use Cases}
\label{subsec:initialusecases}
\autoref{tab:usecaseoverview} presents the outcomes of the content analysis carried out as described in \autoref{subsec:usecaseidentification}. From the twelve interviews conducted, a series of challenges and recurring topics emerged. The table contains short titles to summarise the main challenge as well as a short description or question the local governments seek to find an answer to using digital twin technologies. The identified use cases span a range of impact sectors, including traffic management (cases 6--9 and 14), climate change (cases 11--13 and 15) and wastewater and drainage management (cases 4 and 10). Additionally, the challenges associated with a growing and ageing population are evident in cases 3 and 5, while cases 1 and 2 indirectly relate to high energy consumption and its consequences. Not surprisingly, these impact sectors often overlap; for instance, wastewater management has implications for climate change, while population growth requires a multifaceted governmental response, extending from traffic management and medical services to childcare and housing.

\subsection{Finding Suitable Use Cases}
\label{subsec:summarytable}
\autoref{tab:usecaseprioritisation} presents the matrix used for the internal use case prioritisation. Each of the fifteen cases identified after the content analysis was assessed across multiple variables as previously defined, including usefulness, the need for human behaviour and biological system modelling, whether it requires or could possibly have geographical modelling, if 3D modelling is needed, and its general complexity. This structured evaluation allowed for a clear comparison of potential digital twin applications, taking into consideration that the selected use cases should not only align with the project objectives but also with practical constraints. In particular, as already laid out in \autoref{subsec:usecaseprioritisation}, biological system modelling was quickly decided to be an exclusion criterion, ruling out use cases 4, 11 and 15. In light of their relatively low perceived overall usefulness, use cases 1 and 5--7 were also discarded. For similar reasons, we then also chose to dismiss use cases 12 and 13, as they were deemed to only have medium usefulness, leaving the use cases 2, 3, 8--10 and 14 as the final short list.

After assessing the resources available within our project, we set the scope as selecting a total of three use cases for further development. We thereafter rejected use case 10 due to the high degree of domain knowledge it required and further adapted the final short list by combining use cases 8 and 9. This choice was driven by their shared impact sector and the similarity of the required simulations they would require -- both focus on traffic flow dynamics, requiring similar modelling approaches and data inputs. By combining these use cases into a single use case, we are able to jointly address a complex cluster of challenges and showcase the applicability of digital twin solutions supporting a variety of interdependent issues. Finally, use case 2 was disregarded as a digital twin would not provide the necessary visibility to significantly support decision-making. Alternative information systems were deemed more suitable for that particular context, and will most probably be picked up in future research. After this final decision, we were left with the three use cases deemed most suitable for the project -- use cases 3 and 14 as well as a combination of use cases 8 and 9.

\begin{table*}[!ht]
    \centering
    \resizebox{\textwidth}{!}{
    \begin{tabular}{clccccccc} \toprule
        \multicolumn{1}{c}{\rotatebox{60}{\textbf{No.}}} & 
        \multicolumn{1}{c}{\rotatebox{60}{\textbf{Short title}}} & 
        \multicolumn{1}{c}{\rotatebox{60}{\textbf{Usefulness}}} & 
        \multicolumn{1}{c}{\rotatebox{60}{\textbf{\makecell{Human behaviour \\ modelling}}}} & 
        \multicolumn{1}{c}{\rotatebox{60}{\textbf{\makecell{Biological system \\ modelling}}}} & 
        \multicolumn{1}{c}{\rotatebox{60}{\textbf{\makecell{High degree of required \\ domain knowledge}}}} & 
        \multicolumn{1}{c}{\rotatebox{60}{\textbf{\makecell{Geographical \\ modelling}}}} & 
        \multicolumn{1}{c}{\rotatebox{60}{\textbf{3D modelling}}} & 
        \multicolumn{1}{c}{\rotatebox{60}{\textbf{\makecell{General \\ complexity}}}} \\ \midrule
        1 & Opening hours & \TableLow & \TableYes & \TableNo & \TableNo & \TablePossible & \TableNo & \TableHigh \\
        2 & Blackouts and increased energy consumption & \TableHigh & \TableNo & \TableNo & \TableSimplified & \TablePossible & \TableNo & \TableMedium \\
        \rowcolor{lightgray}
        3 & Population growth & \TableHigh & \TableSimplified & \TableNo & \TableSimplified & \TableNecessary & \TableNo & \TableMedium \\
        4 & Wastewater & \TableHigh & \TableNo & \TableYes & \TableYes & \TablePossible & \TableNo & \TableHigh \\
        5 & Medical supply & \TableLow & \TableSimplified & \TableNo & \TableSimplified & \TableNecessary & \TableNo & \TableMedium \\
        6 & Decluttering of traffic signs & \TableLow & \TableYes & \TableNo & \TableYes & \TableNecessary & \TableNecessary & \TableHigh \\
        7 & Winter service management & \TableLow & \TableNo & \TableNo & \TableSimplified & \TableNecessary & \TableNo & \TableLow \\
        \rowcolor{lightgray}
        8 & Added value through bypass projects & \TableHigh & \TableSimplified & \TableNo & \TableSimplified & \TableNecessary & \TableNo & \TableMedium \\
        \rowcolor{lightgray}
        9 & Traffic flow planning & \TableHigh & \TableSimplified & \TableNo & \TableSimplified & \TableNecessary & \TableNo & \TableMedium \\
        10 & Drainage management and sedimentation planning & \TableHigh & \TableNo & \TableNo & \TableYes & \TableNecessary & \TableNecessary & \TableHigh \\
        11 & Climate change effects on fish populations & \TableLow & \TableNo & \TableYes & \TableYes & \TableNecessary & \TableNo & \TableHigh \\
        12 & Climate change and mobility & \TableMedium & \TableYes & \TableNo & \TableYes & \TablePossible & \TableNo & \TableHigh \\
        13 & Climate change effects on city planning & \TableMedium & \TableNo & \TableNo & \TableSimplified & \TableNecessary & \TablePossible & \TableMedium \\
        \rowcolor{lightgray}
        14 & Shared spaces & \TableHigh & \TableYes & \TableNo & \TableSimplified & \TableNecessary & \TableNo & \TableHigh \\
        15 & Climate change effects on wine & \TableHigh & \TableNo & \TableYes & \TableYes & \TablePossible & \TableNo & \TableHigh \\ \bottomrule
        \multicolumn{7}{c}{\small \parbox{0.9\linewidth}{~\\ \textbf{Abbreviations:} \TableHigh\ = yes/high, \TableMedium\ = medium, \TableLow\ = no/low, \TableSimplified\ = simplified, \TablePossible\ = possible, \TableNecessary\ = necessary}} \\
    \end{tabular}
    }
    \caption{Overview of initial use cases and their evaluation according to the criteria in \autoref{subsec:usecaseprioritisation} -- see there for a more detailed explanation of the abbreviations. The use cases that were ultimately selected are highlighted with a grey background; see \autoref{subsec:summarytable} for details on the thought process leading to the selection of these use cases.}
    \label{tab:usecaseprioritisation}
\end{table*}

\subsection{The Three Use Cases}
\label{subsec:selectedusecases}

\textbf{Use case 1: Population growth.} The first use case puts a focus on managing the challenges posed by different scenarios of population growth for local government. Stakeholders were particularly interested in using digital twins to simulate and visualise possible effects of population growth on traffic patterns, housing demand, childcare services and geriatric care, as well as water, power, sewage and other infrastructure concerns. 

The urgency of this use case was deemed high. Implications for governance mostly centred on decision-making regarding long-term planning for social services, infrastructure and traffic, and the relevance with regard to sustainability was also considered to be very strong. A key concern that emerged was the question of data availability – some data sources (such as different scenarios for population growth or an overview over existing infrastructure and services) were already available, but others (e.\,g.\ granular traffic patterns, water/power/sewage flows) were considered to possibly pose problems. 

\textbf{Use case 2: Traffic flow planning.} The question of traffic flow management in small towns (from more day-to-day decisions like road maintenance scheduling to deciding when and where to construct road bypasses) is highly contentious, with the perceived need for reduction in in-town traffic standing in opposition to possibly worsening emissions and creating induced demand. Against this background, stakeholders put a high value on applying digital twin technology to the simulation of different scenarios to manage traffic flows, to allow for more engagement with the public and lead to more fact-based debates. 

Again, the use case urgency was estimated to be high. Governance connections were seen as assisting decision-making on better planning necessary road maintenance and construction work to cause the least impact to traffic flows, but also more long-term issues like decisions for or against bypass constructions. The corresponding questions of emissions or air quality also had clear sustainability impact. Data availability was seen as less of an issue, with most required data (e.\,g.\ traffic, construction works, air quality, emissions) considered to be easy to procure. 

\textbf{Use case 3: Shared spaces.} Finally, the third use case discussed was planning and redesigning streets and squares as shared spaces. In contrast to the other two use cases, which were considered relevant across a wide variety of locations, this was considered to have a slightly more limited set of possible applications, mostly in those small towns with clear town centres or a strong base of retail businesses in central locations. Nonetheless, urgency was considered high, as the problem of “dying” town centres was deemed a very pressing matter across Lower Austria. Stakeholders were particularly keen on using digital twins (possibly including virtual or augmented reality) to help select streets and squares for urban transformation projects, but also to guide specific details of the implementation of such redesigns. 

Similar to the previous use case, connections to traffic patterns were a strong argument from a sustainability viewpoint. Governance implications were mostly related to various aspects of decision-making on both the selection and the detailed planning (such as placement of plants, benches, lampposts) of shared space projects. Data availability was seen as a key concern, with both 3D models of the relevant streets or squares and detailed mobility patterns (not only motorised individual transport, as in the previous use case, but also e.\,g.\ foot traffic) necessary for a more comprehensive simulation of possible effects of different urban transformation projects.

\section{Discussion}
\label{sec:discussion}

This study applied a transdisciplinary process to identify and refine use cases for digital twin solutions in small towns. The approach integrated scientific knowledge with practical insights from practitioners, ensuring relevance and feasibility in real-world urban planning contexts. The discussion highlights the methodological contributions, practical insights and limitations related to the use case selection process.

\subsection{Methodological Contributions}
The transdisciplinary process facilitated a structured yet flexible framework for identifying and narrowing use cases in small towns. By combining scientific knowledge with stakeholder-driven input, the approach ensured that selected use cases were not only theoretically sound but also practically feasible. This methodology can be adapted to other smart city initiatives by employing participatory workshops, stakeholder engagement and iterative validation. The findings reinforce the importance of scenario configuration, spatial adaptability and multi-criteria analysis in digital twin solutions as aspects that are critical for effective decision support.

\subsection{Practical Insights}
The main challenge that emerged from the interviews and the workshop was to balance feasibility and impact in selecting use cases. Stakeholder alignment is crucial, since different actors prioritise diffferent aspects of urban development, highlighting the importance of governance arrangements in setting the digital transformation agenda of cities and towns. The role of digital twins as communication tools (particularly in engaging citizens through visual representation of planning scenarios) was one of the key findings. Moreover, the need for additional data sources (such as mobile network data) emerged as essential for enhancing the accuracy and applicability of digital twin models. Finally, the workshop also emphasised the importance of integrating real-world experiences from municipalities that have implemented similar urban interventions as well as qualitative data (including surveys capturing the perceptions of e.\,g.\ citizens or business proprietors) were identified as necessary to complement quantitative data, ensuring a holistic understanding of the impact of proposed solutions.

\subsection{Limitations}
Despite the robust methodological approach, the study had certain limitations, especially regarding the selection process which was conducted with a limited number of towns and stakeholders, potentially impacting the generalisability of findings for the defined region. Future research should expand the scope by engaging a broader range of municipalities and testing the methodology across diverse urban contexts as well as including citizens and other stakeholders in the workshops. 

\subsection{Key Findings}
The findings emphasise the value of a transdisciplinary approach in selecting and refining digital twin use cases. By bridging the gap between scientific expertise, policy needs and local knowledge, our research design ensures that digital twin solutions are not only technically feasible but also socially and politically viable. The methodology developed here provides a replicable framework for other small towns seeking to develop digital twin technologies for sustainable urban development. Steps to support the development of those solutions should focus on refining data integration methods, expanding stakeholder engagement and developing user-friendly digital twin interfaces to maximise usability and accessibility for policymakers and citizens alike.

\section{Conclusion and Future Work}
\label{sec:conclusion}
In summary, this research provides a structured methodology for identifying digital twin use cases that can be applied in a variety of different urban contexts. For the purpose of this study, the scope was smaller towns and rural areas in Lower Austria. Nevertheless, the transdisciplinary approach can be extended to both larger cities and other urban, suburban or rural settings. By following the framework outlined in this study, researchers and practitioners can assess local needs, impact sectors and challenges to strategically determine where digital twin applications could enhance decision-making processes.

Moreover, this methodology is not limited to digital twins as a technology supporting decision-making; it can be adapted for the implementation of various digital solutions in cities. The adaptability of the framework lies in its capacity to integrate domain-specific requirements, ensuring that the selection criteria for use cases aligns not only with the expertise of the implementation team but also with the technical constraints, data availability and interoperability needs of the chosen solution. This flexibility enables cities to tailor technological deployments to their unique sociotechnical ecosystems, optimising the efficiency and relevance of smart city initiatives.

Future work focusses on developing a minimum viable product (MVP) and implementing digital twin pilots for each identified use case. This will allow practitioners to interact with a functional prototype, visualising urban data in a dynamic and intuitive way. By allowing a variety of scenario visualisations, the digital twin can support the decision-making process, helping these cities explore potential future developments and policy options. A key objective is to demonstrate the value such digital twins can have as a tool for evidence-based policy design and encouraging its adoption by municipalities and local government. This upcoming research requires the team to understand technical infrastructure limitations, data availability and practical applications of the digital twin simulations, which in particular means to select adequate technologies to support the best possible integration of available data as well as suitable visualisation options.

Fully developed city digital twins are complex systems that require extensive amounts of datasets; small municipalities often lack the necessary infrastructure and data for such digital twin development. Therefore, the MVP (though not a faithful virtual representation of the town) must be as close to reality as possible, to ensure representation of urban conditions. An MVP with attention to detail will allow practitioners to explore key functions, visualise data effectively and understand the potential for further refinement and scalability. Where datasets and data availability represent a challenge, historical data will be used. Moreover, when facing a lack of easily available data, we will make use of statistically plausible interpolated or standardised data following international standards; for example, in the population growth use case, we will adhere to typical values and international standards regarding number of schools per population, required bed availability in hospitals, etc.

During the needs assessment workshop, the value of using the digital twin as a communication tool to engage with residents was highlighted. This was a key input for future work, as the MVP must be designed with an interactive and user-friendly interface to enable transparent and effective public engagement. It is important to note that not all digital twins are suitable for citizen engagement; for those where this is feasible, citizen engagement must be considered in the entire design process from the very beginning~\cite{AbdeenShirowzhan2023, RamuBoopalan2022}. Our MVPs will not allow for direct feedback from citizens, but their design and visualisation will provide a user-friendly interface, thus allowing its use not only for decision-makers but also for the general public. This means the MVP can be used for two ends at once: simulating the scenarios to demonstrate the value and usefulness of digital twins to local government officials as well as serving as a communication support tool for local government towards their residents, helping them to better explain how and why they made certain decisions.

\section*{Acknowledgements}
This research was funded by the Gesellschaft für Forschungsförderung Niederösterreich (GFF NÖ) project GLF21-2-010 ``Smart Cities and Digital Twins in Lower Austria''. The financial support by the Gesellschaft für Forschungsförderung Niederösterreich is gratefully acknowledged.

\bibliographystyle{IEEEtran}
\bibliography{Bibliography}

\begin{thebibliography}{10}
\providecommand{\url}[1]{#1}
\csname url@samestyle\endcsname
\providecommand{\newblock}{\relax}
\providecommand{\bibinfo}[2]{#2}
\providecommand{\BIBentrySTDinterwordspacing}{\spaceskip=0pt\relax}
\providecommand{\BIBentryALTinterwordstretchfactor}{4}
\providecommand{\BIBentryALTinterwordspacing}{\spaceskip=\fontdimen2\font plus
\BIBentryALTinterwordstretchfactor\fontdimen3\font minus
  \fontdimen4\font\relax}
\providecommand{\BIBforeignlanguage}[2]{{%
\expandafter\ifx\csname l@#1\endcsname\relax
\typeout{** WARNING: IEEEtran.bst: No hyphenation pattern has been}%
\typeout{** loaded for the language `#1'. Using the pattern for}%
\typeout{** the default language instead.}%
\else
\language=\csname l@#1\endcsname
\fi
#2}}
\providecommand{\BIBdecl}{\relax}
\BIBdecl

\bibitem{CaragliuDelBo2019}
\BIBentryALTinterwordspacing
A.~Caragliu and C.~F. Del~Bo, ``{Smart Innovative Cities: The Impact of Smart
  City Policies on Urban Innovation},'' \emph{Technol. Forecast. Soc. Change},
  vol. 142, pp. 373--383, 2019. [Online]. Available:
  \url{https://doi.org/10.1016/j.techfore.2018.07.022}
\BIBentrySTDinterwordspacing

\bibitem{MergelEdelmann2019}
\BIBentryALTinterwordspacing
I.~Mergel, N.~Edelmann, and N.~Haug, ``{Defining Digital Transformation:
  Results from Expert Interviews},'' \emph{Gov. Inf. Q.}, vol.~36, no.~4, 2019.
  [Online]. Available: \url{https://doi.org/10.1016/j.giq.2019.06.002}
\BIBentrySTDinterwordspacing

\bibitem{EiblTemple2022}
G.~Eibl, L.~Temple, R.~Sellung, S.~Dedovic, A.~Alishani, and C.~Schmidt,
  ``{Towards a Transdisciplinary Evaluation Framework for Mobile Cross-Border
  Government Services},'' in \emph{21st IFIP WG 8.5 International Conference on
  Electronic Government}, ser. EGOV 2022/Lecture Notes in Computer Science,
  vol.\ 13391.\hskip 1em plus 0.5em minus 0.4em\relax Cham: Springer, 2022, pp.
  543--562.

\bibitem{Scholz2020}
\BIBentryALTinterwordspacing
R.~W. Scholz, ``{Transdisciplinarity: Science for and with Society in Light of
  the University's Roles and Functions},'' \emph{Sustain. Sci.}, vol.~15,
  no.~4, pp. 1033--1049, 2020. [Online]. Available:
  \url{https://doi.org/10.1007/s11625-020-00794-x}
\BIBentrySTDinterwordspacing

\bibitem{AlmulhimSharifi2024}
\BIBentryALTinterwordspacing
A.~I. Almulhim, A.~Sharifi, Y.~A. Aina, S.~Ahmad, L.~Mora, W.~Leal~Filho, and
  I.~R. Abubakar, ``{Charting Sustainable Urban Development through a
  Systematic Review of SDG11 Research},'' \emph{Nat. Cities}, vol.~1, no.~10,
  pp. 677--685, 2024. [Online]. Available:
  \url{https://doi.org/10.1038/s44284-024-00117-6}
\BIBentrySTDinterwordspacing

\bibitem{ScholzSteiner2015}
\BIBentryALTinterwordspacing
R.~W. Scholz and G.~Steiner, ``{The Real Type and Ideal Type of
  Transdisciplinary Processes: Part I—Theoretical Foundations},''
  \emph{Sustain. Sci.}, vol.~10, no.~4, pp. 527--544, 2015. [Online].
  Available: \url{https://doi.org/10.1007/s11625-015-0326-4}
\BIBentrySTDinterwordspacing

\bibitem{Scholz2011}
R.~W. Scholz, \emph{{Environmental Literacy in Science and Society: From
  Knowledge to Decisions}}.\hskip 1em plus 0.5em minus 0.4em\relax Cambridge:
  Cambridge University Press, 2011.

\bibitem{VanDerHornMahadevan2021}
\BIBentryALTinterwordspacing
E.~VanDerHorn and S.~Mahadevan, ``{Digital Twin: Generalization,
  Characterization and Implementation},'' \emph{Decis. Support Syst.}, vol.
  145, 2021. [Online]. Available:
  \url{https://doi.org/10.1016/j.dss.2021.113524}
\BIBentrySTDinterwordspacing

\bibitem{ShiPan2023}
\BIBentryALTinterwordspacing
J.~Shi, Z.~Pan, L.~Jiang, and X.~Zhai, ``{An Ontology-Based Methodology to
  Establish City Information Model of Digital Twin City by Merging BIM, GIS and
  IoT},'' \emph{Adv. Eng. Inform.}, vol.~57, 2023. [Online]. Available:
  \url{https://doi.org/10.1016/j.aei.2023.102114}
\BIBentrySTDinterwordspacing

\bibitem{AlamElSaddik2017}
\BIBentryALTinterwordspacing
K.~M. Alam and A.~El~Saddik, ``{C2PS: A Digital Twin Architecture Reference
  Model for the Cloud-Based Cyber-Physical Systems},'' \emph{IEEE Access},
  vol.~5, pp. 2050--2062, 2017. [Online]. Available:
  \url{https://doi.org/10.1109/ACCESS.2017.2657006}
\BIBentrySTDinterwordspacing

\bibitem{MohammadiVimal2020}
\BIBentryALTinterwordspacing
N.~Mohammadi, A.~Vimal, and J.~E. Taylor, ``{Knowledge Discovery in Smart City
  Digital Twins},'' in \emph{Proceedings of the 53rd Hawaii International
  Conference on System Sciences}, ser. HICSS 2020.\hskip 1em plus 0.5em minus
  0.4em\relax Honolulu, HI: University of Hawaiʻi at Mānoa, 2020, pp.
  1656--1664. [Online]. Available: \url{http://hdl.handle.net/10125/63943}
\BIBentrySTDinterwordspacing

\bibitem{VialePereiraTemple2024}
\BIBentryALTinterwordspacing
G.~Viale~Pereira, L.~Temple, and L.~D. Klausner, ``{Transferring Smart City
  Concepts to Smaller Urban and Rural Contexts: A Systematic Literature
  Review},'' in \emph{Proceedings of Ongoing Research, Practitioners, Posters,
  Workshops, and Projects of the International Conference EGOV-CeDEM-ePart
  2024}, ser. EGOV 2024.\hskip 1em plus 0.5em minus 0.4em\relax Aachen: CEUR
  Workshop Proceedings, 2024. [Online]. Available:
  \url{https://ceur-ws.org/Vol-3737/paper55.pdf}
\BIBentrySTDinterwordspacing

\bibitem{Noy2008}
\BIBentryALTinterwordspacing
C.~Noy, ``{Sampling Knowledge: The Hermeneutics of Snowball Sampling in
  Qualitative Research},'' \emph{Int. J. Soc. Res. Methodol.}, vol.~11, no.~4,
  pp. 327--344, 2008. [Online]. Available:
  \url{https://doi.org/10.1080/13645570701401305}
\BIBentrySTDinterwordspacing

\bibitem{Akremi2022}
\BIBentryALTinterwordspacing
L.~Akremi, ``{Stichprobenziehung in der qualitativen Sozialforschung},'' in
  \emph{Handbuch Methoden der empirischen Sozialforschung}, N.~Baur and
  J.~Blasius, Eds.\hskip 1em plus 0.5em minus 0.4em\relax Wiesbaden: Springer
  VS, 2022, pp. 405--424. [Online]. Available:
  \url{https://doi.org/10.1007/978-3-658-37985-8_26}
\BIBentrySTDinterwordspacing

\bibitem{JohnsonScholes2008}
G.~Johnson, K.~Scholes, and R.~Whittington, \emph{{Exploring Corporate
  Strategy}}.\hskip 1em plus 0.5em minus 0.4em\relax Harlow: Financial Times
  Prentice Hall, 2008.

\bibitem{WellerVickers2018}
\BIBentryALTinterwordspacing
S.~C. Weller, B.~Vickers, H.~R. Bernard, A.~M. Blackburn, S.~Borgatti, C.~C.
  Gravlee, and J.~C. Johnson, ``{Open-Ended Interview Questions and
  Saturation},'' \emph{PLOS ONE}, vol.~13, no.~6, 2018. [Online]. Available:
  \url{https://10.1371/journal.pone.0198606}
\BIBentrySTDinterwordspacing

\bibitem{Schumann2018}
\BIBentryALTinterwordspacing
S.~Schumann, ``{Qualitative empirische Sozialforschung},'' in
  \emph{Quantitative und qualitative empirische Forschung: Ein
  Diskussionsbeitrag}.\hskip 1em plus 0.5em minus 0.4em\relax Wiesbaden:
  Springer VS, 2018, pp. 107--146. [Online]. Available:
  \url{https://doi.org/10.1007/978-3-658-17834-5_6}
\BIBentrySTDinterwordspacing

\bibitem{Mayring2022}
{Mayring, Philipp}, \emph{{Qualitative Inhaltsanalyse: Grundlagen und
  Techniken}}.\hskip 1em plus 0.5em minus 0.4em\relax Weinheim: Beltz, 2022.

\bibitem{TrantasPlug2023}
\BIBentryALTinterwordspacing
A.~Trantas, R.~Plug, P.~Pileggi, and E.~Lazovik, ``{Digital Twin Challenges in
  Biodiversity Modelling},'' \emph{Ecol. Inform.}, vol.~78, 2023. [Online].
  Available: \url{https://doi.org/10.1016/j.ecoinf.2023.102357}
\BIBentrySTDinterwordspacing

\bibitem{AbdeenShirowzhan2023}
\BIBentryALTinterwordspacing
F.~N. Abdeen, S.~Shirowzhan, and S.~M.~E. Sepasgozar, ``{Citizen-Centric
  Digital Twin Development with Machine Learning and Interfaces for Maintaining
  Urban Infrastructure},'' \emph{Telemat. Inform.}, vol.~84, 2023. [Online].
  Available: \url{https://doi.org/10.1016/j.tele.2023.102032}
\BIBentrySTDinterwordspacing

\bibitem{RamuBoopalan2022}
\BIBentryALTinterwordspacing
S.~P. Ramu, P.~Boopalan, Q.-V. Pham, P.~K.~R. Maddikunta, T.~Huynh-The,
  M.~Alazab, T.~T. Nguyen, and T.~R. Gadekallu, ``{Federated Learning Enabled
  Digital Twins for Smart Cities: Concepts, Recent Advances, and Future
  Directions},'' \emph{Sustain. Cities Soc.}, vol.~79, 2022. [Online].
  Available: \url{https://doi.org/10.1016/j.scs.2021.103663}
\BIBentrySTDinterwordspacing

\end{thebibliography}

\appendices
\onecolumn
\section{Interview Guide}
\label{anx:interviewguide}

\begin{table*}[th!]
    \begin{minipage}{\textwidth}
    \centering
    \begin{tabular}{p{8cm}p{8cm}} \toprule
    \bf Question/topic & \bf Objective \\ \midrule
    \multicolumn{2}{c}{\textit{Introduction}} \\
    Short description of the project & The interviewee has a holistic overview of the project. \\
    The goal of this interview is to identify societal challenges in the municipality and look for the possibility of being able to pilot a way of addressing this challenge using the data gathered in the municipality to model a digital twin. & The interviewee knows the general goal and topic of the interview. \\
    A digital twin in the context of this project is a digital representation of the municipality that can simulate various scenarios to enable better decision-making. In other words, it allows to simulate “what if” scenarios and therefore gives decision-makers input for deciding on a particular policy. & The interviewee has a general idea about the concept of digital twins. \\
    Information about the confidentially of collected data and collection of written consent for recording the interview. & Compliance with data protection legislation. \\ \midrule
    \multicolumn{2}{c}{\textit{Current challenges and area of responsibility}} \\
    Could you please introduce yourself briefly and tell us about your role in the municipality? & General information about the person and its role has been gathered. \\
    How long have you been working in your current position? & General information about the person and their role has been gathered. \\
    What societal challenges do you currently see in your area of responsibility?
    \begin{itemize}
        \item If no immediate answer, suggest possible areas: tourism, energy, sustainability/agriculture, waste management, risk control, communication and traffic management, water flow, disaster management.
        \item Specifically: planning public transportation, impact of climate change on tree populations/forestry, power grid overloads, flooding response etc. 
    \end{itemize} & Societal challenges within the persons’ area of responsibility are identified. \\
    Are there specific problems or challenges that seem particularly important to you? Why? & Prioritisation within the societal challenges. \\
    Have you used data and technology previously to address these or other challenges, e.\,g.\ using data as evidence for defining a policy? & Previous ways of tackling these societal challenges are identified. \\
    Are there any existing data regarding these challenges? If so, which ones (e.\,g.\ pollution data gathered by sensors, traffic data from cameras/traffic light sensors, public transport timetables/frequency, hotel occupation numbers, etc.)? & Overview of data availability is provided. \\
    What data do you think would be helpful to better understand and address these challenges? & Further data needs are identified. \\ \midrule
    \multicolumn{2}{c}{\textit{Further challenges and data sources}} \\
    Are you aware of other societal challenges in the municipality that do not directly fall under your area but are handled by colleagues or other departments? & Societal challenges beyond their area of responsibility can be investigated. \\
    What data are used there or would be helpful? & Data sources in other areas are identified. \\
    Are there areas or departments with which you regularly collaborate? What challenges are discussed there? & Knowledge about potential overlaps between use cases is gathered. \\
    Which departments or colleagues are responsible for these challenges? & Potential additional interview partners are found. \\
    Have you included various social actors in addressing any societal challenges, such as residents or enterprises in the area? & Interest in and level of current stakeholder interaction and inclusion is determined. \\ \midrule
    \multicolumn{2}{c}{\textit{Future developments}} \\
    What future developments or trends do you see that could influence the municipality in the coming years? & (Near) Future social challenges are noted. \\
    What additional data or technologies do you think could help better address these developments? & Further data or technologies that could be useful in the project are identified. \\
    What do you hope to gain from a digital twin? In which areas do you think it could be particularly useful? & Expectations towards a digital twin are staked out. \\
    Are there specific use cases or scenarios you would like to see simulated? & Prioritisation for building a digital twin is collected. \\ \midrule
    \multicolumn{2}{c}{\textit{Conclusion}} \\
    Do you have any further comments or questions that have not been discussed yet? & Give the interviewee a chance to add information that was not specifically asked for. \\
    Thank the interview partner for their time and valuable contributions. & Ending of the interview. \\ \bottomrule
    \end{tabular}
    \caption{Interview guide used for the stakeholder interviews as described in \autoref{subsec:datacollection}.}
    \label{tab:interviewguide}
    \end{minipage}
\end{table*}

\end{document}